\documentclass[useAMS,usenatbib]{mn2e}
\usepackage[dvips]{graphicx}
\usepackage{amssymb}
\usepackage{txfonts}

\newcommand{\thisgrb}{Swift J164449.3+573451}

\title[Polarimetry of GRB\,110328 / Swift J164449.3+573451 ]{Polarimetry of the transient relativistic jet of  GRB\,110328 / Swift J164449.3+573451}

\author[K.~Wiersema et al.]
{\parbox{\textwidth}{K. Wiersema,$^{1}$\thanks{E-mail: \texttt{kw113@star.le.ac.uk}}
A. J. van der Horst,$^{2}$
A. J. Levan,$^{3}$
N. R. Tanvir,$^{1}$
R. Karjalainen,$^{4}$
A. Kamble,$^{5}$
C. Kouveliotou,$^{6}$
B.~D.~Metzger,$^{7}$
D. M. Russell,$^{8}$
I. Skillen,$^4$ 
R. L. C. Starling,$^{1}$  and
R.~A.~M.~J.~Wijers$^{7}$}\vspace{0.4cm}\\
\parbox{\textwidth}{$^{1}$ University of Leicester, University Road, Leicester LE1 7RH, UK\\
$^{2}$ Universities Space Research Association, NSSTC, Huntsville, AL 35805, USA\\
$^{3}$Department of Physics, University of Warwick, Coventry CV4 7AL, UK\\
$^{4}$Isaac Newton Group of Telescopes, Apartado de Correos 321, E-38700 Santa Cruz de la Palma, Canary Islands, Spain\\
$^{5}$Center for Gravitation and Cosmology, University of Wisconsin-Milwaukee, 1900 E Kenwood Blvd, Milwaukee, WI - 53211, USA\\
$^{6}$Space Science Office, VP62, NASA/Marshall Space Flight Center Huntsville, AL 35812, USA\\
$^{7}$ Department of Astrophysical Sciences, Peyton Hall, Princeton University, Princeton, NJ 08544, USA\\
$^{8}$Astronomical Institute "Anton Pannekoek," University of Amsterdam, 1090 GE Amsterdam, The Netherlands }}

\begin{document}

\date{14 / dec / 2011}

\pagerange{\pageref{firstpage}--\pageref{lastpage}} \pubyear{2011}

\maketitle

\label{firstpage}

\begin{abstract}
We present deep infrared (Ks band) imaging polarimetry and radio (1.4 and 4.8 GHz) polarimetry of the enigmatic transient \thisgrb. This source appears  to be a short lived jet phenomenon in a galaxy at redshift $z = 0.354$, activated by a sudden mass accretion onto the central massive black hole, possibly caused by the tidal disruption of a star. We aim to find evidence for this scenario through linear polarimetry, as linear polarisation is a sensitive probe of jet physics, source geometry and the various mechanisms giving rise to the observed radiation. We find a formal Ks band polarisation measurement of $P_{\rm lin} = 7.4 \pm 3.5$\% (including systematic errors). Our radio observations show continuing brightening of the source, which allows sensitive searches for linear polarisation as a function of time. 
We find no evidence of linear polarisation at radio wavelengths of 1.4 GHz and 4.8 GHz at any epoch, with the most sensitive 3$\sigma$ limits as deep as 2.1\%. 
These upper limits are in agreement with expectations from scenarios in which the radio emission is produced by the interaction of a relativistic jet with a dense circumsource medium. We further 
demonstrate how the polarisation properties can be used to derive properties of the jet in \thisgrb, exploiting the similarities between this source and the afterglows of gamma-ray bursts. 
\end{abstract}


\section{Introduction \label{sec:intro}}
Recently, the Swift satellite triggered on a peculiar new gamma-ray source. While initially thought to be an unusual gamma-ray burst (GRB\,110328A; Cummings et al. 2011), prior detections by the Burst Alert Telescope (BAT) in addition to  X-ray behaviour highly atypical for GRBs (see e.g. Burrows et al.~2011; Levan et al.~2011) showed this source was likely of a different nature.
Localisation of a variable infrared and radio counterpart (Levan et al.~2011; Zauderer et al. 2011) spatially coincident with the nucleus of a galaxy at redshift $z = 0.3543$ (Levan et al. 2011), further motivated a model in which the transient represents a sudden activation of a blazar-like  phenomenon, most likely fed by the tidal disruption of a (main sequence) star 
(\citealp{Bloom,Cannizzo,Krolik}), though it is difficult to rule out a completely novel phenomenon more closely related to standard gamma-ray bursts (\citealp{Quataert,Woosley}).

Whereas the gamma-ray event itself, as well as the early X-ray behaviour of the transient, is not in accord with predictions of tidal disruption events, the general behaviour at later times seems
to match the expectations for these events (\citealp{Bloom}). In this model, the sudden accretion onto the central massive black hole caused by the tidally disrupted star gives rise to a relativistic, beamed outflow 
whose light dominates over the light produced by the stellar disruption (\citealp{Bloom}).
The spectral energy distribution of the transient shows two prominent bumps (\citealp{Bloom, Zauderer, Burrows}), which may be 
explained by the presence of synchrotron and inverse Compton emission processes (\citealp{Bloom,Burrows,Aliu}). Constraints on the source size from radio spectra (\citealp{Zauderer}) imply a relativistic jet as the source of the observed synchrotron emission, in agreement with fits to the source spectral energy distribution (\citealp{Bloom,Zauderer,Burrows}). This emission is likely generated in the shock interaction between the jet and the circumsource medium, in contrast to blazars in which the synchrotron emission is produced within the jet itself. Modelling of the radio emission indicates a start time of the initial event some 4 days prior to the time BAT triggered, in agreement with analysis of BAT data taken before the trigger occurred (\citealp{Burrows,Zauderer}).

The linear polarisation of jet sources gives crucial insight into the acceleration mechanism of the jet, the configuration of the magnetic field responsible for the synchrotron emission, and the overall geometry of the jet with respect to the line of sight. In the case of this source it may enable us to establish the position of \thisgrb\ within
the family of accreting jet sources, in particular the relation with blazars and X-ray binaries.  In this paper we study the linear polarisation of \thisgrb\ using deep Ks band imaging polarimetry and radio polarimetry, frequencies at which dust obscuration has the smallest influence and at which other transient jet sources are thoroughly studied.  

This paper is organised as follows: in Section 2 we detail our observing strategy, data reduction and analysis; in Section 3 we compare the polarimetry with
models for \thisgrb\ and other jet sources; and in Section 4 we present our conclusions.

\section{Observations and data analysis}\label{sec:obs}
\subsection{WHT LIRIS Ks band polarimetry}
Levan et al.~(2011) show that in the optical range of the spectrum the transient is highly extinguished by dust in the host, and large scale variability is therefore only evident
in the near-infrared wavelengths. This makes optical polarimetry of the transient unfeasible. 
Infrared polarimetry at the longest wavelengths (i.e. K band) is preferred, despite the greater difficulty in obtaining polarimetry with sufficiently small errors of faint sources in this wavelength range. 

We acquired deep Ks band polarimetry of the source using the 4.2m William            
Herschel Telescope at La Palma, Spain, starting on 02:13 UT on April 14th 2011. We used the LIRIS       
(Long-slit Intermediate Resolution Infrared Spectrograph; Manchado et al.~2004) instrument in         
imaging polarimetry mode, which utilises a double Wollaston configuration       
(see Oliva 1997) to obtain simultaneous measurements of the polarised flux at angles 0, 90, 45 and 135 degrees - note that LIRIS does not possess a half-wavelength plate at the time the data were acquired.
An aperture mask is used to avoid overlap of the different angle images, resulting in a field of view of $1 \times 4$ arcmin.

We express the polarimetry information in terms of the Stokes vector $\overrightarrow{S} = (Q, U, V, I)$ (see e.g. Chandrasekhar 1950), where the components of this vector have the following 
meaning: $Q$ and $U$ contain the behaviour of the linear polarisation; $V$ the circular polarisation and $I$ is the total source intensity. Generally we will use the normalised Stokes parameters
$Q/I$ and $U/I$ in this paper. The LIRIS setup means that these can be found from
\[
q = Q/I = \frac{(I_0 - I_{90})}{(I_0 + I_{90})} \,\,\,\,\,{\rm and}\,\,\,\,\, u = U/I = \frac{(I_{45} - I_{135})}{(I_{45} + I_{135})}.
\]
This shows clearly the advantages of a double Wollaston design: these Stokes parameters can be determined from one single exposure, 
under equal conditions, which is particularly advantageous at infrared wavelengths where sky conditions can change rapidly. 
Theoretical models are often expressed in terms of the linear polarisation degree (or fraction) $P_{\rm lin}$ and polarisation angle $\theta$. These quantities can be found from the Stokes parameters
through:
\[
P_{\rm lin} = \sqrt{q^2 + u^2} \,\,\,\,\,{\rm and}\,\,\,\,\, \theta = \frac{1}{2}{\rm arctan}(u/q).
\]
Note that the conversion from Stokes parameters to $P_{\rm lin}$ brings with it complications, discussed further below, so wherever possible we will work in Stokes parameter space.

\begin{figure}   
\centerline{\includegraphics[width=8.5cm]{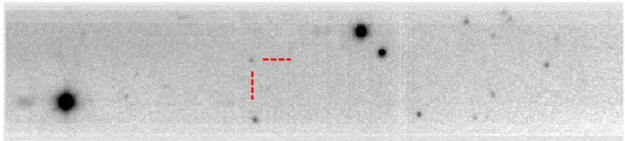}}
\caption{A combined image of a subset of the data of {\it rot1} (-90 degrees position angle), with the transient indicated with tick marks. }
\label{fig:field}
\end{figure}

We obtained a total of 7340 sec  exposure time, using 20 sec integrations and a 5 point dither pattern under     
fair conditions (average Ks band seeing 0.8 arcsec). The dithering was primarily in the mask $X$ direction so the
transient never disappeared behind the aperture mask.
The field orientation, -90 degrees, was chosen such that several bright foreground stars
 fall within the $1 \times 4$ arcmin aperture mask (Fig.~\ref{fig:field}), so that they can serve as secondary calibrators. 
Approximately half the observation, total 3400 sec exposure time, of which three exposures could not be used due to trailing of the stars,    
were taken with -90 degrees rotation. The remainder of the observations were done with field rotated a further 180 degrees
to decrease the effects of flatfielding errors and imperfect behaviour of the Wollaston prisms:
a 180 degree instrument rotation corresponds to a full rotation in $q,u$ space. Hereafter we will call the datasets with the two different rotation angles {\it rot1} and {\it rot2}, respectively.

To calibrate the resulting polarimetry we used observations of zero polarisation standard star BD +33~2642. Observations of this standard consisted of 5 exposures of each 5 sec,
starting at 01:21 UT on April 15th 2011, and were taken without instrument rotation (i.e North up, East left).  

\begin{figure}   
\includegraphics[width=7cm]{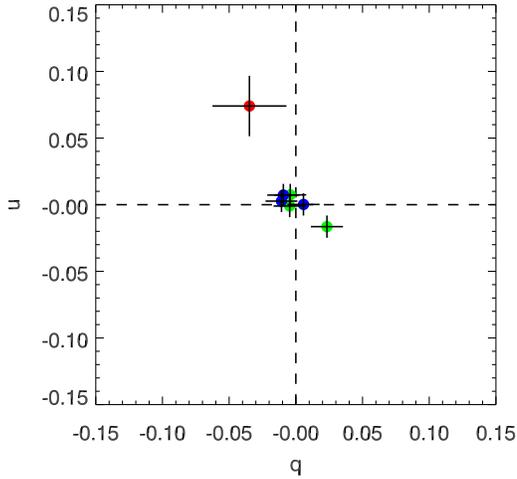}
\caption{This plot shows Stokes parameters $q$ and $u$ of objects in the combined {\it rot1} and {\it rot2}  datasets: bright field stars of the two datasets in green and blue (shifted to the $(q,u) = (0,0)$ point) and \thisgrb\ in red. }
\label{fig:stokes}
\end{figure}

\begin{figure}
\includegraphics[width=7.4cm]{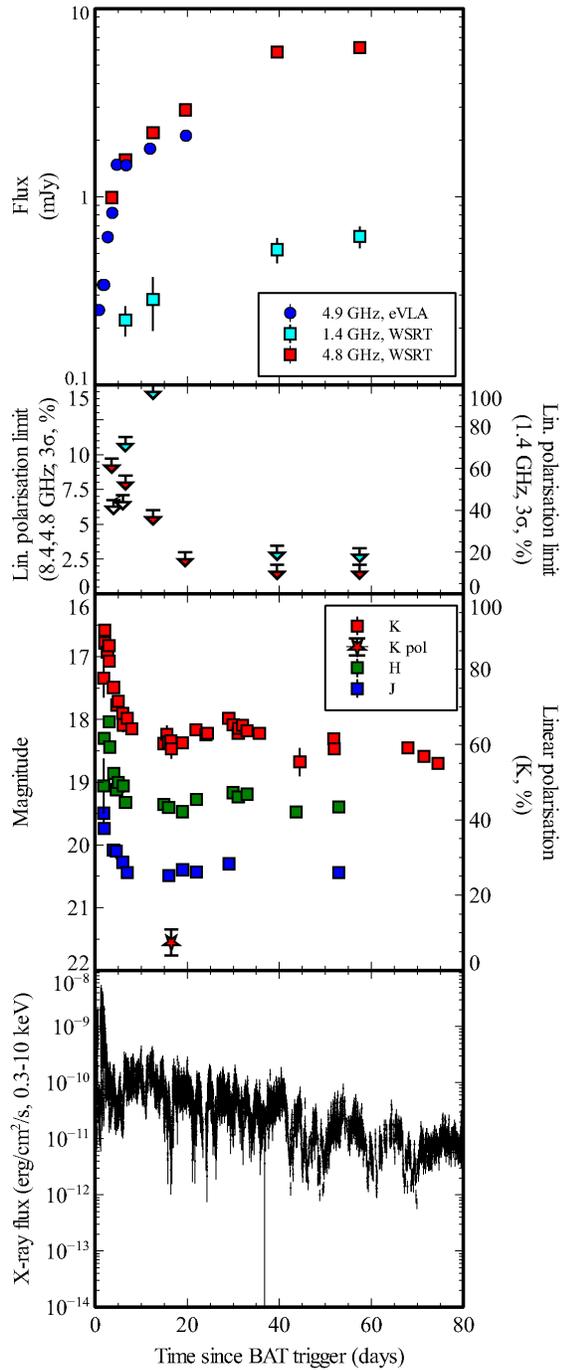}
\caption{The top panel shows our WSRT radio lightcurves at 4.8 GHz and 1.4 GHz (red and cyan, respectively), in combination with EVLA 4.9 GHz data from Zauderer et al. (2011; blue circles).
Note that 
in many cases the errorbars are smaller than the symbol size.
The panel below that shows the 3$\sigma$ limits on the linear polarisation at 4.8 GHz (red, left vertical scale) and 1.4 GHz (cyan, right vertical scale), with added the VLBA 8.4 GHz limits (open symbols, Levan et al.~2011). The third panel shows the 
infrared lightcurves in J, H and K bands (red, green and blue squares, left vertical scale; photometry is not host subtracted), combining data reported in Levan et al.~(2011) and Burrows et al.~(2011) and additional data from Levan et al. in prep. The red star shows the WHT Ks band polarimetry (right vertical scale).  The bottom panel shows the Swift XRT X-ray lightcurve, from the online Swift XRT lightcurve repository (Evans et al. 2007). All lightcurves are plotted with the BAT trigger time (12:57:45 UT on 28 March 2011; Burrows et al. 2011) as zero time.
 }
\label{fig:lcs}
\end{figure}

We clipped from the raw frames the images corresponding to the four different angles, and treat these separately but in an identical way. 
We flat field corrected all data using tasks within IRAF using twilight sky flats taken the same night.  We then performed sky subtraction by constructing sky frames per 
5 dither cycle. The resulting sky subtracted frames were aligned and combined using our own software, separately for the {\it rot1} and {\it rot2} datasets.  
In the same manner we processed the data taken of the zero polarisation standard star.

We used our own software in combination with IRAF routines to perform aperture photometry on the images of the transient to find the source fluxes $I_0$, $I_{90}$, $I_{135}$ and $I_{45}$. We used a seeing matched aperture of 5 pixels radius (1.5 times the on-frame full width at half maximum (FWHM) of the point spread function). A Stokes $I$ image was created as well; photometry performed on this image is shown in the lightcurves in Fig.~\ref{fig:lcs}. It is clear that polarimetry of an object this faint is highly challenging.

We compute the Stokes $q$ and $u$ of the standard star, finding $q = 0.0093 \pm 0.0014$ and $u = -0.0044 \pm 0.0014$. As this source has zero polarisation,
these values can be seen as values for a component of the instrumental polarisation, and the Stokes parameters of the transient (and field stars) can be corrected with these. 
However, we first need to 
bring the observed values for the objects in the {\it rot1} and {\it rot2} data to the same coordinate system as the standard star, through a simple rotation 
\[ 
\left( \begin{array}{c}
q'  \\
u'
 \end{array} \right)
=
\left( \begin{array}{cc}
\cos2\phi & \sin2\phi  \\
-\sin2\phi & \cos2\phi 
 \end{array} \right)
\left( \begin{array}{c}
q  \\
u
 \end{array} \right).
\]
We note that this is not a complete calibration, as the standard and science data have a 90 degree rotation, i.e. the ordinary and extraordinary beams coming out of the Wollaston are swapped,
so we can expect the effects of inefficiencies in the Wollaston (generally speaking of order $P \sim1$\%) to still be present after this calibration. However, the net polarisation
of the transient can be well characterised through the foreground stars in the field of view.

Three bright foreground stars (with $q$ and $u$ uncertainties smaller than 1\%) are detected in both {\it rot1} and {\it rot2} sets, and therefore serve as the most useful 
secondary calibrators. There are several more sources detected in the frames, but we discard them based on proximity to the edges, noticeable asymmetry of the PSF or because they are affected by bad pixels or strong vignetting.
The average of the field star $q$ and $u$ values is somewhat offset from the $(q,u) = (0,0)$ point. This is the combined effect of polarisation induced by Galactic dust scattering (Galactic interstelar polarisation, GIP) and  systematic errors (among others the non-ideal nature of the Wollaston prism, which can in an ideal situation be corrected for by using more instrument rotation angles), as we can assume
that the average net intrinsic polarisation of field stars  is zero. We compute the shift in $q$ and $u$ by taking the average of the field star values, and use the standard deviation as a conservative error.
We then shift the sources in the $q,u$ plane by this amount, and combine the {\it rot1} and {\it rot2} values of the transient by taking the average. To the resulting error we add in quadrature the
uncertainty in the average field star $q,u$ value used for the shift. For the transient we find  $q = -0.035 \pm 0.025$ and $u = 0.074 \pm 0.021$, i.e. Stokes $U$ is non-zero with more than 3$\sigma$
confidence. We plot the resulting $q,u$ diagram in Figure \ref{fig:stokes}. 
From this figure it can be seen that the transient is significantly offset from the field stars, as expected of a truly polarised source, but the uncertainties are substantial, as expected for a source this faint.

Theoretical models are generally expressed in terms of the linear and circular polarisation fractions $P_{\rm lin}$ and $P_{\rm cir}$, and we therefore convert the Stokes parameter information 
to  $P_{\rm lin}$ using the equations above, using the Stokes parameters $q_{\rm cor}, u_{\rm cor}$: the  Stokes parameters obtained when we shift the average of the field stars to $(q,u) = (0,0)$.  
Shifting the distribution by the field star average also eliminates the influence of polarisation induced by GIP, which in any case is expected to be low: 
with $E(B-V)_{\rm Galactic} = 0.02$ (Schlegel et al.~1998) we expect this component of the polarisation to be $<0.5$\%.

The uncertainty on the linear  polarisation angle is a function of the intrinsic polarisation degree, $\sigma_\theta = \sigma_{P_{\rm lin}} / 2P_{\rm lin}$,  i.e. for low polarisation values and faint fluxes the uncertainties in $\theta$ are large. Errors on $q$ and $u$ are generally distributed as a Normal distribution and the Stokes parameters can have positive and negative values. In contrast, $P_{\rm lin}$
is a positive definite quantity, and directly using the equations above will therefore lead to an overestimated $P_{\rm lin}$ and incorrect confidence intervals, an effect generally referred to as the linear polarisation bias.  The correction to the resulting $P_{\rm lin}$ and associated confidence ranges has been studied through both analytical and numerical (Monte Carlo) methods.  Generally speaking, this correction
depends on $\sigma_P$, $P$, and the signal to noise (SNR) of $I_i$.  The Wardle \& Kronberg (1974) prescription is often used in the literature, in which the input $P$ values are multiplied by
$\sqrt{1 - (\sigma_P / P)^2}$ to find the bias-corrected polarisation. We follow Sparks \& Axon (1999) in using a parameter $\eta = P \cdot {\rm SNR}(I_{i})$ to trace the expected behaviour of the bias and $\sigma_P$: when $\eta > 2$ the Wardle \& Kronberg correction is valid, and $\sigma_P$ is as computed directly from the uncertainty of $I_i$. For the transient we find $\eta = 2.3$.

From the relations above we find a formal measurement of the linear polarisation of the transient  in the Ks band  
$P_{\rm lin} = 7.4 \pm 3.5$\% and angle $\theta = -32 \pm 12$ degrees (corrected for polarisation bias). 

\subsection{Radio 4.8 GHz polarimetry}
Radio lightcurves for \thisgrb\ have been reported in Zauderer et al. (2011) and Levan et al. (2011), which can be reproduced by the emission from a relativistic jet.  However, the only reported linear polarimetry of this source  are from two epochs of VLBA observations at 8.4 GHz, giving $2\sigma$ upper limits of 4.5\% and 4.7\% (Levan et al. 2011); we plot these in Fig. \ref{fig:lcs}. 
We report in this paper on radio observations performed with the Westerbork Synthesis Radio Telescope (WSRT) at 1.4 and 4.8~GHz. We used the Multi Frequency Front Ends \citep{tan1991} in combination with the IVC+DZB back end in continuum mode, with a bandwidth of $8 \times 20$ MHz. Gain and phase calibrations were performed with the calibrator 3C~286 at both observing frequencies. The observations have been analysed using the Multichannel Image Reconstruction Image Analysis and Display \citep[MIRIAD;][]{sault1995} software package. 

We observed the source at six epochs, covering April 1 to May 25, of which two observations were taken at 4.8 GHz and the other four at 1.4 and 4.8 GHz. The log and results of these measurements are shown in Table \ref{radioresults}. The flux densities of the first three epochs have also been presented in Levan et al. (2011). Besides the flux densities, we have determined the linear polarisation by generating Stokes $Q$ and $U$ maps. The source was not detected in those maps at all epochs, and Table \ref{radioresults} gives the resulting  $3\sigma$ upper limits on the linear polarisation $P_{\rm lin}$. Due to the low flux density at 1.4~GHz, the limits at this frequency are not well constrained, while upper limits at 4.8~GHz vary from $\sim10\%$ on April 1 down to $\sim2\%$ in the measurements in May 2011. All limits and the resulting lightcurves are shown in Fig. \ref{fig:lcs}. Both 4.8 GHz and 1.4 GHz show a clear, likely achromatic, lightcurve break, which has been clearly seen at higher frequencies in \cite{Zauderer} and discussed in further detail in \cite{Metzger}.

\begin{table}
\begin{center}
\begin{tabular}{|l|l|l|l|l|} 
\hline
Epoch & Duration & Freq. & Flux & $P_{\rm lin}$ \\
(2011) & (hours) & (GHz) & ($\mu$Jy) & ($\%$) \\
\hline\hline
March 31.904 - April 1.393 & 12 & 4.8 & 990$\pm$21 & $<$9.7 \\
April 3.897 - 4.375 & 5.3 & 4.8 & 1573$\pm$28 & $<$8.5 \\
April 9.880 - 10.359 & 5.3 & 4.8 & 2185$\pm$30 & $<$6.0 \\
April 16.861 - 17.349 & 12 & 4.8 & 2893$\pm$20 & $<$3.0 \\
May 6.807 - 7.285 & 5.3 & 4.8 & 5877$\pm$28 & $<$2.1 \\
May 24.757 - 25.236 & 5.3 & 4.8 & 6209$\pm$32 & $<$2.1 \\
\hline
April 3.920 - 4.395 & 5.3 & 1.4 & 221$\pm$43 & $<$75 \\
April 9.903 - 10.379 & 5.3 & 1.4 & 284$\pm$88 & $<$100 \\
May 6.830 - 7.305 & 5.3 & 1.4 & 523$\pm$78 & $<$23 \\
May 24.780 - 25.256 & 5.3 & 1.4 & 614$\pm$77 & $<$22 \\
\hline
\end{tabular}
\caption{WSRT radio flux densities and $3\sigma$ upper limits on the linear polarisation.
\label{radioresults}}
\end{center}
\end{table}

\section{Discussion} 

\subsection{Caveats}
There are a few caveats to point out. Firstly, there is considerable uncertainty on the fraction of the received K band flux that is produced
by the transient source, and what fraction is (stellar) emission from the parent galaxy.
The lightcurve monitoring and optical spectroscopy presented in Levan et al.~(2011) shows that the transient is very red, making the Ks band the most suitable optical/nIR wavelength range to maximise the transient vs.~ host galaxy light ratio. 
It also shows that the host level is likely significantly below the brightness of the source at the time of the polarimetry, which is
further confirmed by late time infrared monitoring which shows that the source continued to fade significantly (at least 0.5 magnitudes) at later times after the 80
day period shown here (Levan et al. in prep.; Fig.~\ref{fig:lcs}).
 In addition, optical spectroscopy shows bright emission lines and a fairly blue host continuum (\citealp{Levan}). As such, we believe that the majority of the light received in the Ks band is from the transient source.

Secondly, the red colour of the transient likely implies considerable dust extinction within the host. Though the exact reddening $A_V$ depends on the assumed model used to fit the spectral energy distribution,
it is likely to be in the range $A_V \sim 3 - 10$ magnitudes (\citealp{Bloom}).  Scattering of light onto dust grains induces polarisation, whose orientation and magnitude as a function of wavelength depends on the geometry of the dust cloud with respect to the emitter and the dust grain size distribution. The high dust extinction values may correspond to a host galaxy dust induced
linear polarisation (HGIP) of a few percent. The HGIP 
should be vectorally added to the intrinsic polarisation (i.e. it adds in $Q,U$ space, so it may increase the measured polarisation $P_{\rm lin}$ or
decrease it, depending on $\theta$).
To derive a crude estimate, we take the parametrisation from \cite{Martin} to describe $p(\lambda)/p_{\rm max}$, take $p_{\rm max} / {\rm E(B-V)} \lesssim 9\%/{\rm mag}$ and assume that all observed Ks band polarisation is caused by dust in the host, we find ${\rm E(B-V) \sim 4}$, which appears too high compared to SED fits. Note that this also implies a polarisation in optical bands a factor 3-4 higher than in the Ks band.
However, it is clearly not possible to quantify the HGIP contribution: we have no information on the dust grain size distribution (as traced to some degree by $R_V$), the exact  extinction value is very uncertain, and the applicability of Serkowski's parametrisation of the relation between extinction and polarisation (Serkowski 1973) is unclear.  A firmer understanding of the extinction within the line of sight in the host galaxy is  required before strong conclusions can be drawn from the absolute value of the presented polarimetry. In particular, polarimetry at multiple broadband filters (or spectropolarimetry)
can in principle differentiate dust scattering geometries (the received infrared light of \thisgrb\ may in fact be an echo, pure scattering) and dependence on grain size distribution (e.g. Zubko \& Laor 2000).

However, we note that the recently discovered source Swift J2058.4+0516 shows remarkable similarity to \thisgrb\ (\citealp{Cenko}), but has very low internal dust extinction, and shows a significant detection of linear polarisation at very similar levels to this source at optical wavelengths (Levan et al. in prep.). 
Based on this similarity we proceed to consider the consequences of the bulk of the detected linear polarisation in the Ks band being intrinsic to the emitting source.

\subsection{The nature of the radio and near-infrared emission and the expected polarisation}
Fits to the spectral energy distribution (SED) and time variable behaviour of \thisgrb\ performed at several different time slices and performed by different groups with independent data (Bloom et al.~2011; Zauderer et al.~2011; Burrows et al.~2011) broadly agree on a number of key points: the SED consists of two prominent bumps, peaking roughly at millimetre and X-ray/gamma-ray frequencies; the X-ray and radio emission are formed in different regions; the presence of a collimated, relativistic jet; the synchrotron nature of the  long wavelength emission.
An acceptable fit to the SED can be achieved with a model where the longer wavelength photons (i.e radio, (sub)mm) are produced by synchrotron emission, and the higher energy photons require the addition of synchrotron self-Compton (SSC) effects or further external Compton (EC) processes (as in blazars). A somewhat poorer fit can be obtained assuming two independent synchrotron components (\citealp{Bloom}; \citealp{Burrows}), requiring a larger extinction at optical wavelengths.  The origin of the inverse Compton seed photons in the two component model is not clear (see e.g. \citealp{Bloom}), but disk photons seem a likely source. The origin of the long wavelength synchrotron photons is more easily diagnosed: the radio-microwave lightcurves and spectra are well reproduced by the emission from the shock produced by the interaction of a relativistic jet with the circumnuclear medium (\citealp{Bloom,Zauderer,Metzger}). The latter paper explores the emission from this shock in more detail, explaining the
achromatic break in the radio light curves (visible in Fig.~\ref{fig:lcs}) by the transition from a phase in which reverse shock emission plays a role in the dynamics, to a phase where a Blandford-McKee self-similar evolution is approached. Our lightcurves show the first detection of this break at 1.4 GHz. 

These models, in which the long wavelength emission comes from an afterglow rather than internal synchrotron production within the jet (i.e. like a blazar; \citealp{vanvelzen,Miller}), make it possible to use inferences from modelling of GRB blastwaves to understand the expected polarisation properties. GRB afterglow polarimetry at optical wavelengths generally shows low but variable linear polarisation levels (below a few percent), and no indications of a significant circular polarisation (\citealp{radiocirc}; Wiersema et al. in prep.). The observed variability in linear polarisation is often seen to track variations in afterglow brightness on top of the general power law decay (most dramatically seen in GRB\,030329, \citealp{Greiner}). This variability is often explained in a scenario where the received emission from a surface of equal arrival time (GRB blastwaves are highly relativistic) consists of a large number of patches, each with a fairly coherent field and high polarisation. As the received emission is the sum over many such patches, the observed linear polarisation is low (\citealp{Gruzinov}). In the case of \thisgrb, an observed polarisation of $\sim7\%$ would imply $\sim100$ such patches (assuming a $\sim70\%$ polarisation of synchrotron emission with a perfectly ordered field).
Bumps in the GRB afterglow lightcurve are formed when distinct bright patches are momentarily dominating the received emission, thereby 
dominating the received polarisation too.  Radio polarimetry of GRB afterglows is highly challenging because of the low fluxes, but predictions indicate detectable levels of polarisation (\citealp{Toma}),
with similar polarisation percentages to the optical polarisation (but note that nearly all GRB afterglows are too faint at radio wavelengths to perform polarimetry at percent level). The non detection of radio polarisation at any epoch in the later time WSRT data is in agreement with the expectation from models explaining the source SED and lightcurves with an afterglow-like phenomenon.  The upper limits from WSRT can be used to probe the physics of the jets in detail once more of the microphysical parameters can be derived (see e.g.  \citealp{Toma} for details).

\cite{Bloom} explicitly indicate that the expected linear polarisation of this source should be relatively low, i.e. like a GRB afterglow. As explained above, we confirm this prediction at radio wavelengths. The modelling of \cite{Metzger} of a larger dataset using this scenario invokes the presence of a reverse shock: the early rise in the radio light curve is occurring while the reverse shock is still passing through the initial shell of ejecta.  However, similar to GRB afterglows, the emission is dominated the foward shock (i.e. the reverse shock only affects the forward shock emission through the dynamics of the jet expansion). Reverse shock emission is occasionally detectable in GRB afterglows (but appears to be fairly rare; \citealp{Klotz,Melandri,Rykoff}), and may be highly polarised, in contrast to forward shock emission, for example in a scenario where there is both a large scale, weak, ordered magnetic field in the ambient medium and a tangled, random, field generated by postshock turbulence (e.g. \citealp{Granot}). In this case the polarisation $P$ varies as a result of changes in the ratio of the ordered-to-random mean-squared field amplitudes (\citealp{Granot}). This model was put forward by the authors to explain the absence of large variation in $P$ and a rotation of $\theta$ at the time of the so-called jet-break, expected if GRB afterglow emission is synchrotron emission dominated by  a tangled magnetic field viewed somewhat off-axis (e.g. \citealp{Lazzati} and references therein). 
We detect no polarisation at radio wavelengths during the steep rise phase in the \thisgrb\  lightcurves with fairly deep limits (Fig. \ref{fig:lcs}), which agrees with the findings of \cite{Metzger} that the received emission is dominated by forward shock emission.
The non detection of any radio polarisation at any epoch also excludes large amplitude variability of $P$, and is consistent with the absence of a jet-break like phenomenon (or other geometric, symmetry breaking effects) under the assumption of a pure tangled magnetic field. Note that \cite{Metzger} infer that $\theta_{\rm jet} \lesssim 1/\Gamma$ throughout: the entire jet is observed, similar to GRBs observed long after the jet break. Further detailed modelling of the jet dynamics will be required to determine the influence of this configuration on the predicted polarisation signal.

Interestingly, the infrared wavelengths find themselves in the region where there are contributions both from the synchrotron component seen in radio wavelengths, and from the second bump in the SED, where inverse Compton effects play an important role. The detection of a low level of polarisation, described above, is not inconsistent with the expectations from an afterglow origin of the long wavelength emisison, and is comparable with GRB afterglows (see Wiersema et al. in prep. for Ks band polarimetry of the afterglow of GRB\,091018), though 7\% polarisation is high for a GRB afterglow.
This value is in line with the non-detection of radio polarisation in this scenario (\citealp{Toma}). 
However, there is a possibility for a contribution of non-synchrotron emission processes to the measured polarisation. The expected polarisation properties of inverse Compton emission have been evaluated by \cite{Kraw} in the framework of blazar emission, who demonstrates that SSC emission from unpolarised seed photons produces vanishing linear polarisation for electron Lorentz factors $\gamma \gtrsim 10$. While the origin of the seed photons for this source is not well established, it seems likely that they originate from somewhere in the accretion disk (\citealp{Bloom}) or are the same photons that form the radio-optical spectrum (see \citealp{Aliu} for a discussion), so in both cases are effectively unpolarised. The analytical and numerical results of \cite{Kraw} indicate that it is unlikely that SSC contributes to the observed Ks band polarisation.

We can now contrast the polarisation properties of \thisgrb\ with other transient relativistic jet sources, in particular X-ray binaries. In these sources the detected synchrotron emission is internal to the jet, not from an afterglow like phenomenon. As such, their polarisation properties are different in some aspects from GRB afterglows. Radio polarimetry of low mass X-ray binary jets shows the polarisation and its position angle to be a function of wavelength and spectrum. For optically thick jets (flat spectrum, self-absorbed)  the synchrotron polarisation is low (a few percent), whereas for optically thin jets it is much higher 
(10 - 30\%; see \citealp{Fender} for a review). In the optical and near-infrared wavelengths there may be a contribution from Rayleigh scattering in the vicinity of the accretion disk (the massive star plays a role as well in high mass 
X-ray binaries), and when jets are seen these wavelengths diagnose the magnetic field closer to the base of the jet (e.g. \citealp{Russell}). Sometimes this is optically thin and the polarisation is then seen to be a few percent and highly variable (e.g. \citealp{Russell}).  
Of particular interest for comparison with \thisgrb\ is the low mass X-ray binary (black hole candidate) XTE J1550-564: during a faint X-ray outburst in the low-hard state, Ks band polarimetry was obtained by \cite{Dubus}, 
who find an intrinsic (i.e. after correcting for the Galactic dust induced polarisation through field stars) linear polarisation in the Ks band of 0.9\% - 2\% (95 \% confidence interval). The authors attribute this 
polarisation to optically thick synchrotron emission from a compact jet of the binary. Their measurements of low level polarisation in this case, far from the self-absorption frequency, is comparable to the polarisation of \thisgrb. When the origin of the near-infrared emission of \thisgrb\ is better understood (through long timescale lightcurves and late time SEDs), the comparison with XTE J1550-564 and X-ray binaries in general can be further quantified.

Finally, the comparison of \thisgrb\ with blazars has been made several times in the literature. As also remarked in \cite{Levan}, the polarisation of this source is markedly lower than sometimes observed in                                                                            blazars (\citealp{Aller}) but it is generally consistent with what is seen when they are in a relatively low flux state: the most strongly optical / near-infrared polarised emission appears to correspond to the peak of blazar flares. Our LIRIS observations of \thisgrb\ occurred at relatively late times, and the evolution of the event was much more rapid than typical blazar flares.  Multi-epoch polarimetry of events like \thisgrb\ when they are closer to the peak brightness may therefore shed more light on the link of this source with blazar polarisation variability.

As a last point we wish to point out the suitability of sources like \thisgrb\  for X-ray polarimetry missions, for example GEMS, which would have been able to perform percent-level polarimetry in a reasonable exposure time ($\sim100$ ks) at a few days after trigger (see Fig.~\ref{fig:lcs} for the X-ray lightcurve).
 The X-ray and radio brightness of \thisgrb\ and its transience (i.e. tracking the synchrotron break frequencies) allows for new insights into the physics of jet launching, particle acceleration and photon emission processes: X-ray polarimetry can distinguish between external Compton and SSC effects, while ground-based (radio - optical) studies can determine the micro- and macrophysics of the jet-medium interaction.

\section{Conclusions}\label{sec:conclusions}
We present in this paper a measurement of the Ks band linear polarisation of \thisgrb, giving Stokes parameters $Q/I = -0.035 \pm 0.025$ and $U/I = 0.074 \pm 0.021$, or  
$P_{\rm lin} = 7.4 \pm 3.5$\%. In addition we present an extensive set of upper limits on the linear polarisation at two
radio frequencies, 4.8 and 1.4 GHz.  These measurements confirm the results from SED modelling that the long wavelength emission originates in the interaction of a relativistic jet with the medium around the black hole in the centre of the host galaxy. The SSC component inferred from SED modelling is not likely to contribute to the observed polarisation in the Ks band. It seems likely that a polarisation component caused by HGIP is present in the Ks band polarimetry. It would require a better knowledge of the shape of the extinction curve and the amount of extinction $A_K$ to correct for this. However, the recent detection of
Swift\,2058.4+0516 shows that relatively unobscured versions of \thisgrb\ do exist, where HGIP is negligible. In those less obscured cases multi-colour polarimetry can be obtained, which allows for 
HGIP correction in a direct way.
Our radio polarimetry restricts the possible configuration of the magnetic field and its coherence through limits on polarisation during both the steep rise phase in the radio lightcurves and deep limits at late times.   
Lastly, our study demonstrates the possibility to probe these very faint transients using 4m class telescopes (with highly versatile instruments like LIRIS). 

\section*{Acknowledgments}
We thank the ING staff for their support of the LIRIS polarimetry effort  discussed here, in particular M. Hrudkova. We thank the anonymous referee for useful suggestions. KW acknowledges support from STFC. RLCS is supported by a Royal Society Fellowship.
The William Herschel Telescope is operated on the island of La Palma by the Isaac Newton Group in the Spanish Observatorio del Roque de los Muchachos of the Instituto de Astrof'sica de Canarias.
 The WSRT is operated by ASTRON (Netherlands Institute for Radio Astronomy) with support from the Netherlands foundation for Scientific Research. This work made use of data supplied by the UK Swift Science Data Centre at the University of Leicester.

\label{lastpage}

\end{document}